\documentclass[aps,prl,reprint,groupedaddress,showpacs]{revtex4-1}

\usepackage[]{graphicx}

\begin{document}


\title{Generation of isolated asymmetric umbilics in light's polarization}


\author{Enrique J. Galvez}\email[]{egalvez@colgate.edu}  \author{Brett L. Rojec}
%
\affiliation{Department of Physics and Astronomy, Colgate University, Hamilton, NY, U.S.A.}
\author{Vijay Kumar and Nirmal K. Viswanathan}
\affiliation{School of Physics, University of Hyderabad, Hyderabad, India}


\date{\today}

\begin{abstract}
Polarization-singularity C-points, a form of line singularities, are the vectorial counterparts of the optical vortices of spatial modes and fundamental optical features of polarization-spatial modes. Their generation in  tailored beams has been limited to lemon and star C-points that contain symmetric dislocations in state-of-polarization patterns. In this article we present the theory and laboratory measurements of two complementary methods to generate isolated asymmetric C-points in tailored beams, of which symmetric lemons and stars are limiting cases; and we report on the generation of monstars, an asymmetric C-point with characteristics of both lemons and stars.

\end{abstract}


\maketitle




Line singularity patterns provide an important way to characterize topologies and their curvatures. Ridge patterns in fingerprints provide a vivid display of line singularities \cite{Penrose}. More often, line singularities are abstract characterizations of the principal curvatures of gaussian topological features \cite{BerryHannay,Beuman}. Applications include shape interrogation and face recognition \cite{Halinan},  diagnosis of vector fields such as those exhibited by the polarization of the sky \cite{BerrySky}, or in the ellipse patterns of speckle fields \cite{SoskinOL03,Flossmann,Egorov,Flossmann2}. 

So far, the study of line singularities relies on the diagnosis of natural occurrences. Non-separable superpositions of polarization and spatial mode of light can provide a vehicle for deliberately creating line-singularity patterns for their study. This polarization-spatial-mode hybridization also adds a new dimension to imaging, where polarization provides additional sensing information \cite{Tyo,OPN}. Different species exploit polarization-spatial combinations for their survival \cite{Brady}, and line singularities may provide the means to characterize them. At the quantum level, these hybrid modes provide larger Hilbert spaces for encoding information \cite{Nagali}. An investigation of polarization-spatial light modes is also essential for understanding this type of imaging at a deeper level \cite{Milione}.

C-points are the umbilical points of the line singularities in the polarization of light because they  connect the apex of two opposite cones (a diabolo): of semi-major and semi-minor axis lengths \cite{BerryHannay,Umbilics}. They consist of a state of circular polarization surrounded by a field of polarization ellipses in the optical field, with orientations that rotate about the C-point  \cite{NyeHajnal}. C-points are singular points of ellipse orientation. They are intimately linked to the optical vortices of scalar fields, but encode the optical dislocations in the state of polarization instead of the phase \cite{BerryDennis01}. The production of the full spectrum of C-points is of interest in its own right, as it reveals a new domain of complex light not investigated before. 
The production and analysis of C-points are the basis for new techniques to produce and diagnose optical vortices, which are of interest in metrology due to their high sensitivity to perturbations \cite{Loffler}. 

The two symmetric types of C-point singularities, known as lemons and stars, correspond to dislocations where the ellipse orientation varies linearly with and counter to the angle about the singularity, respectively. 
Yet, the generation of beams bearing isolated C-points (i.e., alone in a light beam) has been limited to these two cases \cite{Beckley,GalvezAO,Marrucci}. The larger class of asymmetric C-points containing orientations evolving nonlinearly have been produced only in speckle patterns \cite{SoskinOL03,Flossmann,Egorov,Flossmann2}, or as C-point pairs (dipoles) in tailored beams \cite{KumarJO13}.

The two symmetric cases are the ends of a spectrum of C-points where the pattern of orientations in the ellipse field is nonlinear and asymmetric. Within asymmetric C-points is a hybrid type of C-point, the monstar, which has features of both lemons and stars. Monstars have been predicted theoretically \cite{BerryHannay,DennisOC02}, but have not been produced as isolated singularities.  
In this article we present the theoretical framework and two experimental arrangements to generate and analyze the full range of isolated C-points, including monstars. We do so via complementary studies in our two laboratories.



\begin{figure}[htb]
\hfil\scalebox{1}{\includegraphics{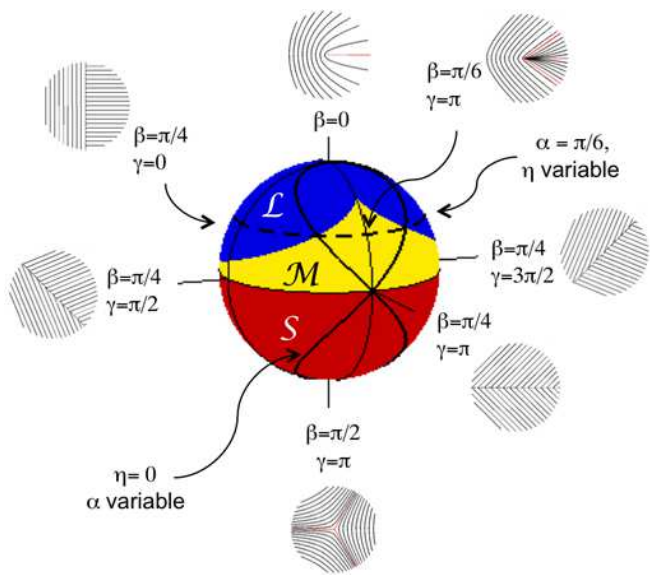}}\hfil 
\caption{C-point sphere denoting by color the three morphological regions of C-points: lemons (L), monstars (M) and stars (S). The polar coordinates are $2\beta$ and $\gamma$ (for $\delta=0$), as specified by Eq.~\ref{eq:monstar}. Inserts show polarization lines corresponding to selected points on the sphere. }
\label{fig:monstardom}
\end{figure}

C-points appear in the optical field when we combine an optical vortex in one state of circular polarization with a plane wave in the opposite state of polarization \cite{DennisOC02,BerryDennis01}. 
If we represent the polarization field of a C-point in polar coordinates $(r,\phi)$ by
\begin{equation}
\Psi=(\cos\beta\; re^{{\rm i}\phi}+\sin\beta\; re^{{\rm -i}\phi}e^{{\rm i}\gamma})e^{{\rm i}\delta}\hat{e}_R+\hat{e}_L,
\label{eq:monstar}
\end{equation}
where $\hat{e}_R$ and $\hat{e}_L$ denote states of right and left circular polarization, respectively, we can map all types of C-points onto a unit sphere, proposed here, and shown in Fig.~\ref{fig:monstardom}. 
The basis for all asymmetric C-points is an asymmetric vortex (first term), formed by a superposition of vortices of opposite topological charge. 

In the topology of Gaussian surfaces, C-points are umbilical points of degenerate curvature \cite{BerryHannay}. Line-singularity patterns of light, known as ``polarization lines,''  connect the semi-major axes of the polarization ellipses.   The polarization lines of a lemon contain one angular direction where all the ellipse's semi-major axes are radial, as shown in  Fig.~\ref{fig:monstardom} for the (symmetric) case $\beta=0$. Star C-points have three angles with radial polarization lines, as shown in Fig.~\ref{fig:monstardom} for the case $\beta=\pi/2$.

As the pattern of polarization lines is made asymmetric, a third type of umbilic appears: the monstar. Similar to the lemon, this type of C-point has ellipse orientations that rotate in the same sense as the angular coordinate $\phi$; and similar to the star, it has three angles where the polarization lines are radial. 
The monstar region of Fig.~\ref{fig:monstardom}, ``monstardom,'' has zero width for $\gamma=0$. It has a maximum north-ward extension in a cusp-like termination of the region for $\gamma=\pi$. Starting from Eq.~\ref{eq:monstar}, at a given point $(r,\phi)$ the orientation of the ellipse $\theta$ is half of the relative phase between the two polarization components, or $\theta=\phi$. After some algebra the latter becomes a cubic equation in  $\tan\phi$. 
The  case $\gamma=\pi$ can be solved analytically: the point at the tip of the cusp corresponds to $\tan\beta=1/3$ \cite{GalvezSPIE13}. 
The equatorial line represents a unique set of modes that  involve $\pi$-phase shear singularities carrying C-lines of circular polarization but no C-points \cite{FreundOC02}. 

Varying the phase $\delta$ in Eq.~\ref{eq:monstar} does not produce new patterns: a C-point with parameters $(\beta,\gamma,\delta)$ is the same as the one with $(\beta,\gamma-2\delta,0)$ but rotated by $\delta$.   The normalized areas  of the regions on the sphere are: 0.382 for lemon, 0.118 for monstar and 0.5 for star. With proper weighting \cite{DennisOL08}, they give the correct density fractions of C-points found in random fields: 0.447, 0.053 and 0.5, respectively \cite{Flossmann2,DennisOC02,DennisOL08}.

The sphere is fundamentally related to 
the sphere of first-order spatial modes proposed earlier \cite{CourtialPadgett}, where any mode is a superposition of antipodal modes. 
This way, we can represent a C-point as a superposition of any two C-point antipodes on the  sphere. In this work we create the C-points experimentally via two sets of antipodes: Polar antipodes (i.e., Eq.~\ref{eq:monstar}) via superpositions of Laguerre-Gauss modes, and equatorial antipodes (below) via superposition of Hermite-Gauss modes.

In the approach using antipodal equatorial modes, $x$ and $y$ are the spatial basis:
\begin{equation}
\Psi=(\cos\psi\; x+\sin\psi\; ye^{i\alpha})e^{i\eta}\hat{e}_R+\hat{e}_L.
\label{eq:monstarHG}
\end{equation}
The spatial mode for the case $\alpha=0$, $\psi=\pi/4$ and $\eta=0$ is implemented with a Hermite-Gauss mode: HG$_{10}$ rotated by 45 degrees, which is proportional to $(x+y)$ times a gaussian function; the latter does not affect the morphology of the C-point. (The same is true with Laguerre-Gauss modes for implmenting the optical vortices of Eq.~\ref{eq:monstar}.) Its antipode, at $\alpha=\pi$, $\psi=\pi/4$ and $\eta=0$ is the mode HG$_{10}$ rotated by $-45$ degrees.  The correspondence between the two sets of angles can be obtained after some algebra. A simple case adapted to the experimental conditions keeps $\psi=\pi/4$. For this case,
$\beta=\pi/4-\alpha/2$, 
$\gamma=-\pi/2$, and 
$\delta=\eta+\alpha/2-\pi/4$. The thick solid and dashed lines on the sphere correspond to the sequence of states followed when varying $\alpha$ with $\eta=0$, and varying $\eta$ with $\alpha=\pi/6$, respectively.

\begin{figure}[htb]
\hfil\scalebox{1}{\includegraphics{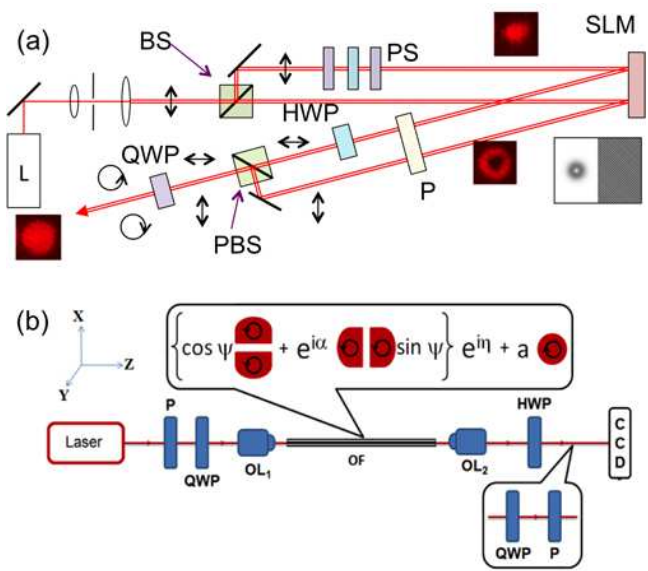}}\hfil 
\caption{Apparatuses to produce C-points: (a) using Laguerre-Gauss spatial modes, which includes a helium-neon laser (L), beam expander with spatial filter, non-polarizing (BS) and polarizing (PBS) beam splitters, half-wave plate (HWP), quarter-wave plate (QWP), Pancharatnam-Berry phase shifter (PS), polarizer (P), and spatial light modulator (SLM). Insert shows an example of the programming of the SLM; (b) using Hermite-Gauss spatial modes, with the following additional components: objective lenses (OL) and optical fiber (OF).}
\label{fig:app}
\end{figure}

We present two laboratory approaches to preparing tailored beams bearing isolated C-points. In one approach (done at Colgate U.), we implemented Eq.~\ref{eq:monstar} with the polarization interferometer of Fig.~\ref{fig:app}a. A spatially-filtered vertically-polarized optical beam from a helium-neon laser entered the interferometer via a non-polarizing beam splitter. One of the arms was phase shifted by $\delta$ using a Pancharatnam-Berry phase shifter. The two beams were incident onto a spatial light modulator (SLM), which encoded in half pane, a phase-blazed amplitude-modulated superposition of Laguerre-Gauss modes. The mode appeared diffracted  0.5 degrees form the specular reflection. The other pane had a plain blazed grating.  A half-wave plate in the path of one of the beams rotated its linear polarization to  horizontal. A mirror and a polarizing beam splitter recombined the two beams. A quarter-wave plate placed after the interferometer put the two polarization components in circular polarization states.

We diagnosed the polarization pattern of the light using imaging polarimetry, which involved gathering six images obtained via polarizing filters. These images were used to find the Stokes parameters for each pixel of the imaged beam, and thus the complete state of polarization \cite{GalvezSPIE13,KumarJO13}.


In the other approach (done at U. Hyderabad), we performed superpositions of  antipodal  equatorial modes (Eq.~\ref{eq:monstarHG}), implemented by superpositions of Hermite-Gauss modes. This was done via  intra-optical-fiber manipulation of polarization and spatial modes, as shown in Fig.~\ref{fig:app}b. A circularly-polarized Gaussian beam was launched skew off-axially into a step-index circular-core two-mode optical fiber (V \#: 3.805, 37.4-cm long) by means of an objective lens (0.4 NA). This excited a superposition of a circularly polarized anisotropic vortex and an orthogonally polarized fundamental mode. Thus, the fiber-mode dynamics generated isolated C-points on their own, as shown in previous studies  \cite{Jayasurya,Kumar,Kumar2}. The launch angle of the light into the fiber selected angles $\alpha$ and $\psi$ in the implementation Eq.~\ref{eq:monstarHG}.
The light emerging from the fiber was collimated and the relative phase between the two circular components ($\eta$ in Eq.~\ref{eq:monstarHG}) was adjusted by a half-wave plate.  The detection method was the same as the one described earlier.

\begin{figure}[htb]
\hfil\scalebox{1}{\includegraphics{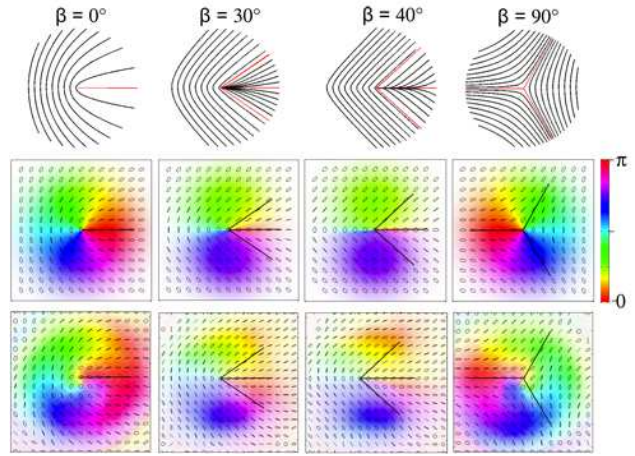}}\hfil 
\caption{Results of theoretical calculations and experimental measurements when using Laguerre-Gauss modes as the spatial mode basis (Eq.~\ref{eq:monstar}). The first row has the calculated polarization lines, with the radial lines shown in red. The second and third rows show the modeled and measured ellipse fields. The solid lines in the ellipse fields correspond to the predicted orientations of the radial lines. The false color represents the orientation of the ellipses, and the saturation of the color is proportional to the intensity of the beams. }
\label{fig:LGmonstars}
\end{figure}

The results obtained with the apparatus of Fig.~\ref{fig:app}a are shown in Fig.~\ref{fig:LGmonstars}. The first row has the predicted polarization lines, the second row has the theoretically predicted ellipse field, and the third row has the measured images, with the ellipse field obtained from the polarimetric analysis. The false color indicates the orientation of the ellipses, which is also the  Stokes field \cite{SoskinOL03, FreundOC02}, defined as $\arg(S_1+{\rm i}S_2)$, where $S_1$ and $S_2$ are the Stokes parameters. 
The saturation of the color is proportional to the intensity of the light in the beam.

The first and last columns of Fig.~\ref{fig:LGmonstars} show lemon and star C-point fields. The right-handed circularly polarized points are surrounded by a right-handed ellipse-field region bounded by a circular ``L'' line of linearly polarized states.
At radii larger than the L-line, the states of polarization are left-handed. 
We added the predicted radial polarization lines (0 for $\beta=0$; 0, $\pm 34.2^\circ$ for $\beta=30^\circ$; 0, $\pm 42.2^\circ$ for $\beta=40^\circ$; and $180^\circ, \pm 60^\circ$ for $\beta=90^\circ$) to guide the eye. 

The middle columns show two cases involving monstars. As can be seen, the ellipse patterns show unambiguously the characteristics of monstars. The measured orientation of the ellipses agrees with the expectations. The experimental patterns also show characteristics that are common in laboratory conditions, such as slight misalignments and Gouy-phases. These can be seen in the curved boundaries between colors, which are especially pronounced for $\beta=0$ and $\beta=90^\circ$.  The theoretical maps do not model these effects.

\begin{figure}[htb]
\hfil\scalebox{1}{\includegraphics{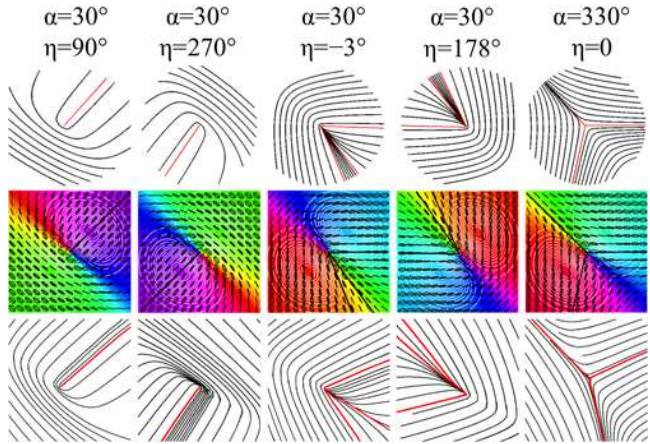}}\hfil 
\caption{Results of data taken using the Hermite-Gauss spatial basis modes (Eq.~\ref{eq:monstarHG}). The  first and third rows are the calculated and measured polarization-line maps, with the radial lines shown in red. The second row shows the modeled ellipse fields, with solid lines corresponding to the predicted orientations of the radial lines, white contours denoting intensity contours, and colors denoting ellipse orientation.}
\label{fig:HGmonstars}
\end{figure}

The results using Hermite-Gauss spatial modes are shown in Fig.~\ref{fig:HGmonstars}. The first row has the simulated polarization lines for each case, the second row has the modeled pattern, and the third row shows the measured polarization-line maps \cite{Flossmann,Flossmann2}, which are the polarization lines extracted from the data. The background colors in the second row represent the orientation of the ellipses in the field.
All C-points shown have asymmetric patterns. 
The first and second columns show lemons obtained with parameters $\alpha=30^\circ$ and $\eta=90^\circ$, and $\alpha=30^\circ$ and $\eta=270^\circ$, respectively. They are equivalent to those produced via Eq.~\ref{eq:monstar} with $\beta=30^\circ$, $\gamma=90^\circ$ and $\delta=60^\circ$; and $\beta=30^\circ$, $\gamma=90^\circ$ and $\delta=240^\circ$, respectively. In contrast to the symmetric lemon of Fig.~\ref{fig:LGmonstars} (case $\beta=0$), these lemons are highly asymmetic. 
The fifth column of Fig.~\ref{fig:HGmonstars} is an asymmetric star with $\alpha=330^\circ$ and $\eta=0^\circ$.  C-points with $\alpha=330^\circ$, are stars  with nearly equatorial latitude on the sphere, equivalent to a value of $\beta$ of $60^\circ$. 

The third and fourth columns of Fig.~\ref{fig:HGmonstars} are highly asymmetric monstars with $\alpha=30^\circ$ and $\eta=-3^\circ$; and $\alpha=30^\circ$ and  $\eta=178^\circ$, respectively.
These cases are near the lemon-monstar boundary, and where there is radial-line degeneracy. For example, when $\eta=180^\circ$ (and $\alpha=30^\circ$), two radial lines appear at the same angle: $\phi=150^\circ$, with the third one appearing at  $\phi=180^\circ$. 
As $\eta$ is increased, the radial line disappears creating a lemon with $\phi=180^\circ$. Conversely, if $\eta$ is decreased, the radial line splits forming a monstar, as shown, with radial lines at 178.7$^\circ$, 125.5$^\circ$, and 113.9$^\circ$. The measured patterns agree with the expectations. Discrepancies in the location of the radial lines are due to errors in the fiber launch angles for setting $\alpha$ and $\psi=45^\circ$.

In summary, we have described two methods to prepare optical beams bearing asymmetric C-points. We have presented experimental demonstrations of the production of these asymmetric polarization singularities, which include monstars, an asymmetric C-point not produced previously in isolation. In addition, we have presented that the space containing asymmetric singularities can be mapped onto the two-dimensional surface of a sphere.

Polarization-spatial modes hold promise for the discovery of new phenomena and applications. When we add polarization to scalar fields, the modal landscape changes dramatically: singular points may no longer be dark, and interferometry techniques yield to polarimetry \cite{FreundOL01}. The use of of light beams with amplitude and polarization modulations may allow a finer type of manipulation where the forces depend on the field gradient and on the orientation or state of the electric field relative to the objects' axis of symmetry \cite{Ambrosio}. At the quantum-mechanical level, the asymmetric C-points that we produce are a spatial mode qutrit ``entangled'' with the polarization qubit of a single photon. The merging of spatial modes and polarization  promises advances in imaging and information multiplexing with light. It opens new possibilities for understanding the ways in which these manifest in nature \cite{Brady}, an area that has yet to receive much attention. More generally, light polarization provides a framework for a deeper study of line singularities and their use to describe natural phenomena.

\begin{acknowledgments}
E.J.G and B.L.R. acknowledge support from the National Science Foundation, U.S. Air Force contract FA8750-11-2-0034, and Research Corporation. V.K. and N.K.V. thank the Department of Science and Technology and Council of Scientific and Industrial Research for financial support. We thank K. Beach, M.V. Berry, X. Cheng, M.R Dennis, F. Flossmann and G. Milione for help and stimulating discussions. 
\end{acknowledgments}


\end{document}